\documentclass[twocolumn,english,aps,floatfix,superscriptaddress]{revtex4}
\usepackage[T1]{fontenc}
\usepackage[latin9]{inputenc}
\usepackage{amsmath}
\usepackage{graphicx}
\usepackage{esint}

\makeatletter
\@ifundefined{textcolor}{}
{%
 \definecolor{BLACK}{gray}{0}
 \definecolor{WHITE}{gray}{1}
 \definecolor{RED}{rgb}{1,0,0}
 \definecolor{GREEN}{rgb}{0,1,0}
 \definecolor{BLUE}{rgb}{0,0,1}
 \definecolor{CYAN}{cmyk}{1,0,0,0}
 \definecolor{MAGENTA}{cmyk}{0,1,0,0}
 \definecolor{YELLOW}{cmyk}{0,0,1,0}
}

\usepackage{graphics}

\@ifundefined{definecolor}{\@ifundefined{definecolor}
 {\usepackage{color}}{}
}{}

\usepackage{bm}

\usepackage{babel}

\usepackage{babel}

\usepackage{babel}

\usepackage{babel}

\@ifundefined{definecolor}{\@ifundefined{definecolor}
 {\@ifundefined{definecolor}
 {\@ifundefined{definecolor}
 {
}{}
}{}
}{}
}{}

\usepackage{babel}

\usepackage{babel}

\makeatother

\usepackage{babel}

\newcommand{\bS}{{\bf S}}
\newcommand{\br}{{\bf r}}
\newcommand{\bk}{{\bf k}}

\newcommand{\be}{\begin{equation}}
\newcommand{\ee}{\end{equation}}

\begin{document}

\title{Odd-frequency Two Particle Bose-Einstein Condensate}

\author{ A.V. Balatsky}

\date{\today}

\affiliation{NORDITA, KTH Royal Institute of Technology and
Stockholm University,Roslagstullsbacken 23
106 91 Stockholm, Sweden}

\affiliation{Institute for Materials Science, Los Alamos National Laboratory,
Los Alamos, NM 87545, USA}

\begin{abstract}
We introduce the concept  of the {\em odd-frequency} Bose Einstein Condensate (BEC), characterized by the  odd frequency/time two-boson expectation value.   To illustrate the concept of odd frequency BEC we present simple classification of pair boson condensates that explicitly permits this state. We point qualitative differences of odd-frequency BEC with conventional BEC and introduce the order parameter and wave function for the odd-frequency BEC.
\end{abstract}
\maketitle

{\em Introduction}  Bose-Einstein condensate (BEC) is defined as a state where ground state attains the macroscopic finite expectation value of a single particle boson field $b_a(\br)$ \cite{BEC1,BEC2,BEC3}.  The general case scenario so far has been described as
\begin{equation}\label{EQ:BEC1}
\langle b_{a}({\bf r}) \rangle  = b_0 \neq 0
\end{equation}
for a boson annihilation operator of state $a$, that can be  encompassing  both spin, orbital index, band or flavor. For simplicity we will  assume  $a, b = 1,2$ being orbital labels. From this basic definiton the mactroscopic expectation value of a single particle density matrix follows as well $  \langle b^{\dag}_a({\bf r}) b_b({\bf r'})\rangle =  |b_0|^2 $ as $ |{\bf r} - {\bf r'}| \rightarrow \infty $. At no point in this discussion one needs to invoke the time dependence of operators and of expectation values.  The condensate means the steady $b_0$ and hence time is an irrelevant variable. Numerous experiments with He4 and with cold atoms firmly establishes the phenomenon of BEC with macroscopic expectation value of a single boson field \cite{BEC4,BEC5,BEC6}.

There are cases when single particle expectation value is zero yet there is a nontrivial order in the system. Typical example would be  spin nematic, e.g. \cite{Nematic1,Nematic2, Gorkov90, Podolsky05,Momoi12} . Classification of magnetically ordered states starts with the single spin expectations
 \be \langle \bf{S}(\bf{r}) \rangle \neq 0 \ee.
 If the single spin expectations are zero this does not mean that the state is paramagnetic as nematic order is still possible. The order parameter for the magnetic nematic is a two spin correlator
 \be \langle \bS(\br) \rangle = 0,  \langle S^i(\br) S^j(\br')\rangle  = Q^{ij}(\br,\br') = Q(n^in^j- 1/3 \delta_{ij})\ee
  with nontrivial spin structure captured by nematic vector $\bf{n}$ is still possible.   Moreover if one is allowed to keep track of time dependence in two spin correlator one can have even and odd-time magnetic correlations present in the spin system. The former corresponds to the conventional spin nematic state.  Later ones was shown to be related to the chiral spin states \cite{Nematicodd}.
  Similarly one classifies possible symmetries of the BEC by exploring higher order correlations even in cases when the single boson expectation is trivially zero $  b_0 = 0$.

Purpose of this paper is to  introduce  a class of BECs that is qualitatively different from the  conventional BEC and involves nontrivial, i.e. {\em odd in time} correlations, we call it odd frequency two particle
 BEC. It also can be viewed as an odd frequency BEC nematic. We also will use interchangeably the term odd-frequency BEC, having in mind Matsubara representation the odd in time and odd frequency orders are identical.
To make the claim we have to  broaden the discussion by explicitly considering i) two boson condensate similar to the spin nematic case and ii) we need to keep track of time dependence of $b_{a}({\bf r}, \tau)$. \cite{Matsubara}. We consider
\be
\label{EQ:Def1}
 D_{ab}({\br- \br'}|\tau - \tau') = \langle T_{\tau} b_a({\br},\tau) b_b({\br'}, \tau') \rangle
 \ee
where $D_{ab}(\br,\tau)$ is a function of relative coordinate and time for a translationally invariant and equilibrium states  assumed here. $T_{\tau}$ stands for Matsubara time ordering. It is possible to have  a nontrivial BEC with $b_0 =0$. Namely when two boson condensation occurs one would have
\begin{equation}\label{EQ:BEC2}
  \langle b_{a}({\bf r}, \tau = 0) \rangle  = 0, D_{ab}(\br - \br'|\tau - \tau')\neq 0
\end{equation}
with $D_{ab}(\br|\tau )$ being the "nematic"  version of the boson order.   The notion that the zero time expectation value is zero would correctly imply that there is no single particle condensate yet it would be wrong to infer that there is no condensate. One can have a condensate of a higher order based on the fact that $D_{ab}(\br,\tau)$ is a nontrivial function. This function can be short ranged function of $\br - \br'$ yet have finite value as a function of center of mass coordinate $1/2(\br + \br')$.  One naturally would conclude that there is a long range order of pairs of bosons. Global $U(1)$ gauge symmetry is exlicitly broken in the state with finite $D_{ab}$ and it describes the macroscopically coherent state:
\be
\label{EQ:U1}
b_a \rightarrow \exp(i\theta)b_a; D_{ab} \rightarrow \exp(2i\theta) D_{ab}.
\ee

 The anomalous function   $D_{ab}$ can be either an {\em even} or   an {\em odd} function of $\tau$ \cite{comment1}.  If it is an {\em even} function of time we can consider its value at $\tau = 0$ as an order parameter and we find that it corresponds to a staightforward two boson condensate. Example of this condensate would be a BEC formed of two particle molecules. If $D_{ab}$ is an {\em odd} function of time $\tau$ then we are looking at the state that has clearly broken symmetry yet has no simple equal time correlator  so we can not use as  equal time correlator an order parameter as $D_{ab}(\br|\tau  =0) =0$.
  One has to define the order parameter of this coherent state differently.    Thus we point the existence of  the class of two boson condensates that are bosonic analog of odd-frequency superconducting condensate. \cite{Ber,BA} Symmetry proof of  existence of the class of odd frequency boson nematic or two boson BEC is  the main result of this paper.

{\em Classification} We start with classification of symmetry  of boson condensates.
Define orbital permutation operator $O$ as $O D_{ab} = D_{ba}$.  Bose statistics determines that simultaneous permutation of time ( T reversal) , coordinate inversion (P) and permutation of orbitals  should leave $D$ invariant $D_{ab}(\br, \tau) = D_{ba}(-\br, -\tau)$ \cite{Nematicodd}.   We can write it simply as
\be
\label{EQ:Def2}
P T O D_{ab}(\br, \tau) = D_{ab}(\br, \tau)
\ee
 shorthand as $PTO = +1$.  Thus we can define the table of all possible parities with respect to discrete symmetries discussed here:
\\

  \begin{tabular}{|c|c|c|c|}
    \hline
     & P & T & O \\
     \hline
    even-$\omega$ & $+$ & $+$ & $+$ \\
    \hline
     even-$\omega$ & $-$ & $+$ & $-$ \\
     \hline
     odd-$\omega$ & $-$ & $-$ & $+$ \\
     \hline
    odd-$\omega$ & $+$ & $-$ & $-$ \\
    \hline
   \end{tabular}
  \label{EQ:Tab1}\\
  \\
{\em Table 1. Description of separate discrete properties of each of the operations $P,T, O$. The pair of states labeled odd-$\omega$ are the odd frequency states that satisfy basic condition $PTO = 1$.}

There are therefore multiple condensates one can observe. For the even$-\omega$ condensates we have states that will have only even frequency harmonics and they can be represented by equal time correlators that are nonzero.  They are what we call even frequency states. For example one can develop an P even T even and diagonal in orbital index order that would correspond to a on-site two boson condensate. This case, while interesting on its own, is not the focus of our discussion.

The alternative possible state that follows from this table is the odd-$\omega$ states that are also allowed. We will focus now on these so called odd-frequency states.

 {\em Odd-frequency  two boson BEC}

There is no restriction on the function $D_{ab}(\br | \tau)$ other than the symmetry of the condensed state.  The odd-time BEC also implies that this function is odd in (Matsubara) frequency: $D_{ab}(\br, \tau)$ can be expanded in Fourier harmonics and one can see that only odd-frequency harmonics will contribute. The detailed form of $D_{ab}$ will be set by interactions. For simplicity consider the linear term $D_{ab}(\br, \tau) = d_{ab}(\br) \tau$ at small $\tau$. Then the {\em first} nontrivial expectation function at $\tau = 0$ is the derivative
\begin{equation}\label{EQ:BEC3}
\partial_{\tau} \langle b_{a}({\bf r}, \tau)b_b(0,0)\rangle |_{\tau = 0} = \partial_{\tau} D_{ab}(\br, \tau)|_{\tau = 0}  = d_{ab}(\br).
\end{equation}

$d_{ab}(\br)$ shall be used as an {\em order parameter} for the {\em odd frequency} or {\em odd time}  two particle BEC. We retain band (species, flavor) index $a,b = 1,2$ explicitly. Qualitatively same results follow  for a single boson species when one can ignore $ a $ dependence.

The function $D_{ab}(\br, \tau)$ that is odd in time is a more complicated object than $d_{ab}(\br)$ as it contains other information aside from its derivative at zero time. Yet this simple  order parameter does capture the odd nature of the condensate and is a minimal statement one can make about the order parameter. The  situation here is similar to the  Ginzburg-Landau (GL) order parameter   vs the full microscopic theory with the frequency dependence of the anomalous functions. In the GL sense one simply is dealing with the constant order parameter and ignores time/frequency dependence, i.e. retardation.

 {\em Example} To evaluate the time derivative on $b$ one needs to take a commutator with the Hamiltonian. We  introduce  a specific model to illustrate that our definition of the order parameter corresponds to composite condensate. We present few examples of how the odd frequency condensate and related order parameter would appear in microscopic theory. The example is meant  to illustrate the consistency of definitions  and not meant to be a definitive microscopic theory of composite BEC that still needs to be developed.

 i) Holstein coupling of $b$ to the classical field $x$:
 \begin{eqnarray}\label{Eq:BEC4}
   H = \sum_{a,{\br}} \omega_0 b_a^{\dag}(\br)b_a(\br) + g_a(b_a(\br)+b_a^{\dag}(\br)) x(\br) + \nonumber \\ kx(\br)^2/2 + m\dot{x}^2(\br)/2.
 \end{eqnarray}
For the displacement $x$ we assume that the equilibrium position is at $\langle x\rangle_{g_a =0} =0 $, evaluated for the nonperturbed Hamiltonian of displacive interactions. Both $b_a$ and $x$ are fields that depend on coordinate $\bf r$.
We have
 \begin{equation}\label{Eq:BEC5}
   \partial_{\tau} \langle b_{a}({\bf r}, \tau)b_b(0,0) \rangle |_{\tau = 0}  = d_{ab}(\br)  \propto g_a\langle x(\br,0)  b_b(0,0)\rangle_{\tau = 0}.
 \end{equation}
 where we used equations of motion  $\partial_{\tau} b_a \sim \omega_0 b_a + g_a x$. The {\em composite}  condensate of $\langle x b_a \rangle$, while neither x nor b itself are ordered,$\langle x \rangle = \langle b_a \rangle = 0$  is  the order parameter of odd-frequency BEC in $\langle b_a b_b \rangle$.

 For completeness, we also discuss the wave function for this odd frequency condensate. As in the case for odd frequency supersonductors one might ask the question about the wave function that describes the condensate \cite{Dahal,Abrahams}. In the steady state the  wave function  would contain no time dependence even for odd frequency condensate. We use the order parameter for odd frequency two particle BEC: $d_{ab(\br)}  = g_a \langle x(\br) b_a(\br)\rangle $. The natural  choice   is the polaron like wave function that describes the coherent state with macroscopic condensation of composite boson $ \partial_{\tau} b_a({\bf r}, \tau) ~ \hat{x}({\bf r}) $. It takes the form
 \be
 \label{EQ:WF1}
 |\Psi\rangle \sim \exp[ \sum_{a, \br} (\lambda_a({\bf r})b^\dag_a(\br) \hat{x}({\bf r}) - \hat{x}^2({\bf r})/2\sigma)]|0\rangle
 \ee
The first term  generates the coherent states of the $ x b_a({\bf r})$ and second term is introduced to keep the expectation value $\langle x({\bf r})\rangle $ finite; $\lambda_a$ and $\sigma$ are the variational parameters that would determine conditions for the formation of such composite condensate. Direct evaluation shows  $ \langle \Psi| b_a({\bf r})x({\bf r})|\Psi\rangle \sim \lambda_a({\bf r})\langle 0|x({\bf r})|^2| 0\rangle $, where $\lambda_a({\bf r})$ being the condensate wave function (explict position dependence would be a consequence of the external confining potential.  Thus we verify that odd-frequency two particle BEC is related to conventional composite BEC $\langle xb_a\rangle$.

Using these wave functions we  can prove that the composite condensate and odd frequency BEC  forms  a class, orthogonal to conventional BECs with same quantum numbers of the condensates. The  wave function of conventional BEC is  $|\Psi_0 \rangle \sim \exp[\sum_{\bf r} \nu_a({\bf r}) b^{\dag}_a({\bf r})]|0 \rangle $. By direct inspection one finds that $\langle \Psi_0|\Psi \rangle  = 0$; conventional single boson BEC  is orthogonal to the state in Eq.(\ref{EQ:WF1}), as long as  $\langle \Psi| x ({\bf r})  |\Psi \rangle = 0$. This  example supports the view that proposed {\em odd frequency} BEC is a novel kind of Bose condensate $~\langle x b_a \rangle$ and without need to have a single boson condensate.

So far the claims we made are a consequence of the symmetries. The logic to proceed would be same as in conventional BEC: functions  $ \lambda_a({\bf r})$ can be obtained from solving microscopic theory using Eq.(11) as a variational wave function. Any model with nonzero $\lambda_a$ in the ground state would support composite BEC as a ground state phenomenon.  More precise statements will be made once we have a detailed microscopic model that generates the composite condensate.

 ii)  two layer BEC. Consider the case where the band index $a, b$ representing the layer index that bosons can occupy. The Hamiltonian is:
 \begin{eqnarray}
 H = \sum_{a,b,\bk} \varepsilon_a(\bk) b^{\dag}_{a,\bk}b_{a,\bk} + \sum_{a,b,\br}V_{ab}n_a(\br)n_b(\br) +\nonumber \\
  t(b^{\dag}_{a}(\br)b_b(\br) + h.c.)
 \end{eqnarray}
\label{EQ:bilayer1}
with boson boson interaction $V_{ab}$ and interlayer tunneling $t$.  Now, assuming there is no conventional BEC in each layer: $\langle b_a \rangle = 0$, consider the case of odd-frequency intralayer BEC. Then the nonzero two particle correlation is
  diagonal in layer index and odd in time $D_{aa}(\br|\tau)$, as in Eq(\ref{EQ:BEC3}). Assuming linear term in $\tau = 0$ suffice we get (keeping only relevant terms):
\be
\label{EQ:Bilayer2}
\partial_{\tau} b_a({\bf r}, \tau) \sim t b_b(\br, \tau)
\ee
and the order parameter of the odd frequency in layer BEC is
\be
\label{EQ:Bilayer3}
d_{aa}(\br) \sim t \langle b_a(\br) b_b(\br) \rangle .
\ee
The odd-frequency {\em intraleyer} two particle BEC is equivalent to the interlayer two particle even frequency BEC.
The wave function corresponding to the odd-frequency BEC in this case will be a simple coherent state or paired bosons on both layers $\Psi \sim \exp(\lambda \sum_{\br} b^{\dag}_a(\br)b^{\dag}_b(\br))$. This  example  demonstrates the consistency of the formalism of odd frequency two particle BEC and relates its order parameter to conventional two boson condensate. The situation here resembles the case of multiband superconductors where direct interband hybridization  induces the odd-frequency superconducting component \cite{ABSB}, for a review see \cite{Tanakarev}. Here the interlayer tunneling $t$ also plays the role of interband hybridization and allows nonzero odd-frequency BEC.

We  comment on the  feasibility to realize this odd frequency two boson BEC.
The composite nature of the condensate in case of odd frequency BEC makes it difficult to realize. The fact that there is a multiparticle condensate would mean restrictions additional constraints on the phase space that need to be satisfied. One finds it is  difficult to create odd ferquency two boson condensate or, equivalent,  composite bose condensate without creating single particle condensate. One natural route to the creation of composite condensate would be BEC proximity effects.  In the presence of the media that possess extra low energy excitations  the conventional BEC that will tunnel will get dressed by these low energy excitations and will develop odd frequency component. Interlayer coupling is one such example. We are in the process of developing this model.
Another obvious next step is the search for microscopic models that allow the formation of odd-frequency BEC in cold atoms BEC  with multiple bosonic species.   The tunability of interactions and ability to create multiple species with tunable interactions would make this these systems a promosing candidate for odd frequency pair BEC realization. At present we do not have a model that supports two bosong BEC as a ground state. We therefore have to limit ourselves here to the purely symmetry discussion and  leave questions of a microscopic model for a separate discussion.

{\em Conclusion}
We point the existence of a new class of two boson condensate that is an odd-frequency BEC or odd-frequency nematic BEC. This new class of boson nematic condensate adds to the growing list of other odd-frequency states like odd frequency superconductors \cite{Ber} and odd-frequency spin nematic \cite{Nematicodd}. Various  odd-frequency states point to the existence of whole parallel realm of largely unexplored odd-frequency many body ordered states in addition to conventional even-frequency states. Promising candidate models that could support odd-frequency BEC nematic would be  proximate structures where interband or interlayer hybridization and tunneling induced this novel condensate. Investigation of realistic  microscopic models would further elucidate the conditions for existence and nature of of odd-frequency BEC.

{\em Acknowledgements} Discussions with E. Abrahams, R. Barnett, A. Black-Schaffer Y. Kedem, F. Mancarella are gratefully acknowledged.
Work at NORDITA was supported by  VR 621-2012-2983 and ERC-DM-321031. Work  at Los Alamos was supported by the Office of Basic Energy Sciences. Los Alamos National Laboratory, an affirmative action equal opportunity employer, is operated by Los Alamos National Security, LLC, for the National Nuclear Security Administration of the U.S. Department of Energy under contract DE-AC52-06NA25396.


\begin{thebibliography}{10}

\bibitem{BEC1}
Bose, S. N.  "Plancks Gesetz und Lichtquantenhypothese". Zeitschrift für Physik {\bf 26},  178, (1924). doi:10.1007/BF01327326.
\bibitem{BEC2} Einstein, A.  "Quantentheorie des einatomigen idealen Gases". Sitzungsberichte der Preussischen Akademie der Wissenschaften {\bf 1},  3.(1925).
    \bibitem{BEC3} Landau, L. D.  "The theory of Superfluity of Helium 111". J. Phys. USSR {\bf 5},  71, (1941).
L. Landau . "Theory of the Superfluidity of Helium II". Physical Review {\bf 60} (4),  356 (1941). doi:10.1103/PhysRev.60.356.

\bibitem{BEC4} M.H. Anderson, J.R. Ensher, M.R. Matthews, C.E. Wieman, and E.A. Cornell, "Observation of Bose Einstein Condensation in a Dilute Atomic Vapor". Science {\bf 269} (5221), 198, (1995).  doi:10.1126/science.269.5221.198.

\bibitem{BEC5} C. C. Bradley, C. A. Sackett, J. J. Tollett, and R. G. Hulet,  "Evidence of Bose Einstein Condensation in an Atomic Gas with Attractive Interactions". Physical Review Letters {\bf 75} (9), 1687 (1995).  doi:10.1103/PhysRevLett.75.1687.

\bibitem{BEC6}  K.B. Davis, et al., Phys. Rev. Lett. {\bf 75} (1995) 3969.

\bibitem{Nematic1} A.F. Andreev and I.A. Grishchuk, Sov. Phys. JETP 60,
267 (1984) [Zh. Eksp. Teor. Fiz. {\bf 87}, 467 (1984)].

\bibitem{Nematic2} A. V. Chubukov, Phys. Rev. {\bf B 44}, 4693 (1991).

\bibitem{Gorkov90} L. Gorkov and A. Sokol, JETP Lett, {\bf 52}, 504, (1990).

\bibitem{Podolsky05} D. Podolsky and E. Demler, New J. Phys, {\bf 7}, 59, (2005).

\bibitem{Momoi12} T. Momoi et al, Phys. Rev. Lett., {\bf 108}, 057206, (2012).

\bibitem{Nematicodd} A.V. Balatsky and E. Abrahams, Phys. Rev. Lett. {\bf 74}, 1004, (1995).

\bibitem{Matsubara} The Matsubara time $\tau$ is defined on a finite range $ 0 \leq \tau \leq \beta$ , $ \beta = 1/T$. The Bose statistics dictates that $b_{a}({\bf r}, \tau)$ is  a  periodic function of $ \tau \rightarrow \tau  + \beta $.

\bibitem{comment1} Simple inspection shows that being odd in $\tau$ and periodic in $\beta$ does pose some constraints on the possible functions but there are functions that satisfy both, e.g. $f(\tau) = sin(2\pi \tau/\beta)$.

\bibitem{Ber} Berezinskii V. L.  JETP Lett. {\bf 20} 287, (1974).

\bibitem{BA} Balatsky A and Abrahams E,  Phys. Rev. {\bf B 45}, 13125, (1992).



    \bibitem{Abrahams} Abrahams E, Balatsky A, Scalapino D J and Schrieffer J R,  Phys. Rev. {\bf B 52} 1271, (1995).
    \bibitem{Dahal} Hari P Dahal et al,  Wave function for odd-frequency superconductors,   New J. Phys. {\bf 11} 065005, (2009).

        \bibitem{inhom} We allow for spatially inhomogeneous condensate as is the case in the cold atom traps.

        

        \bibitem{ABSB} Annica M. Black-Schaffer and Alexander V. Balatsky, "Odd-frequency superconducting pairing in multiband superconductors",
Phys. Rev. {\bf B 88}, 104514 (2013).


\bibitem{Tanakarev}  Yukio Tanaka, Masatoshi Sato, and Naoto Nagaosa, "Symmetry and Topology in Superconductors - Odd-Frequency Pairing and Edge States,"J. Phys. Soc. Jpn. {\bf 81}, 011013 (2012)

 \end{thebibliography}
\end{document}